# DEMONSTRATION OF NIOBIUM TIN IN 218 MHz LOW-BETA QUARTER WAVE ACCELERATOR CAVITY*

T.B. Petersen[†], M.P. Kelly, T. Reid, M. Kedzie, B. Guilfoyle, G. Chen
Argonne National Laboratory, Lemont, IL, USA
S. Posen, B. Tennis, G. Eremeev, Fermi National Accelerator Laboratory, Batavia, IL, USA

*Abstract*

A 218 MHz quarter wave niobium cavity has been fabricated for the purpose of demonstrating Nb3Sn technology on a low-beta accelerator cavity. Niobium-tin has been established as a promising next generation SRF material, but development has focused primarily in high-beta elliptical cell cavities. This material has a significantly higher $T_C$ than niobium, allowing for design of higher frequency quarter wave cavities (that are subsequently smaller) as well as for significantly lowered cooling requirements (possibly leading to cryocooler based designs). The fabrication, initial cold testing, and Nb3Sn coating are discussed as well as test plans and details of future applications.

## INTRODUCTION

Niobium-tin (Nb$_3$Sn) has been identified as the most promising next-generation superconducting material for accelerator cavities. The main reason for this choice is the higher critical temperature ($T_C$ = 18.3 K compared to 9.2 K for pure niobium), which corresponds to a significantly lower surface resistance for a given temperature and frequency. This is a consequence of the dependence of the Bardeen-Cooper-Schrieffer (BCS) resistance on material characteristics like critical temperature $T_C$ among others. [1] The relationship between frequency, critical temperature, operating temperature, and $R_{BCS}$ (which is inversely proportional to RF power losses/cryogenic load) is illustrated in Figure 1 below (generated using SRIMP code) [2].

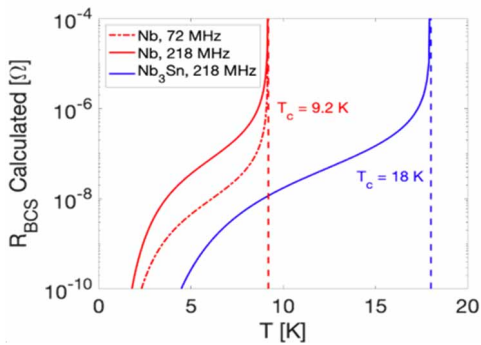

Figure 1. Predicted BCS surface resistance of Nb and Nb3Sn vs operating temperature for two frequencies.

The incentive for utilizing Nb$_3$Sn is two-fold: drastically reduce the size of low-beta cavities for fabrication as well as the operational cooling requirements. This could lead to cryo-cooler based stand-alone cryomodules a fraction of the physical size of previous generations, with the cavity scale shown in Figure 2.

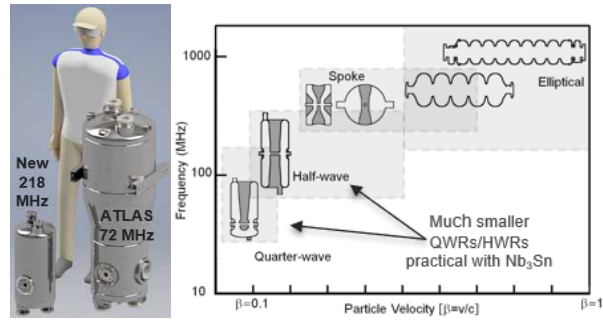

Figure 2. Physical scale of 218 MHz cavity compared to similar geometry 72 MHz cavity in ATLAS. This technology could expand the useful range of different cavity geometries.

## CAVITY DESIGN AND FABRICATION

The cavity design was aimed at demonstrating a useful low-beta cavity near 200 MHz and with a peak magnetic field of 60 mT. The frequency was chosen as a multiple of the ATLAS clock, with the goal of installing two of these 218 MHz cavities in ATLAS. Cavity EM parameters are shown below in Figure 3 [3].

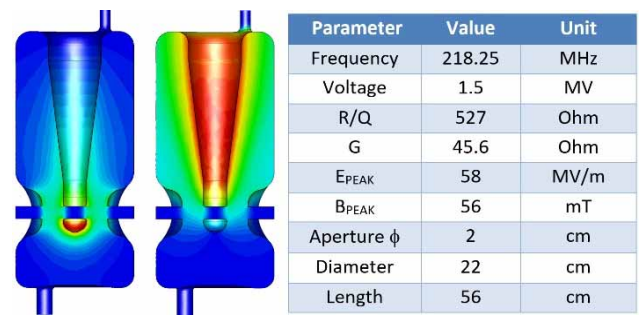

| Parameter | Value | Unit |
| --- | --- | --- |
| Frequency | 218.25 | MHz |
| Voltage | 1.5 | MV |
| R/Q | 527 | Ohm |
| G | 45.6 | Ohm |
| $E_{PEAK}$ | 58 | MV/m |
| $B_{PEAK}$ | 56 | mT |
| Aperture φ | 2 | cm |
| Diameter | 22 | cm |
| Length | 56 | cm |

Figure 3. Cavity field distributions and EM parameters.

*This material is based upon work supported by the U.S. Department of Energy, Office of Science, Office of Nuclear Physics, under contract number DE-AC02-06CH11357, and the Office of High Energy Physics, under contract number DE-AC02-76CH03000. This research used resources of ANL's ATLAS facility, which is a DOE Office of Science User Facility.
[†]tpetersen@anl.gov

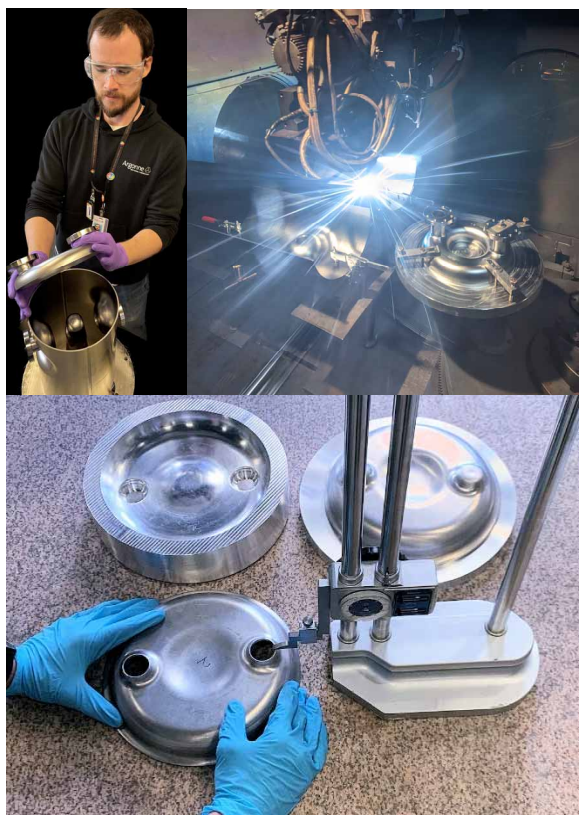

Figure 4. Cavity parts during fabrication (left, bottom) and electron beam welding (right).

The 218 MHz cavity was fabricated from high RRR niobium with hydroforming techniques utilizing a local vendor, Stuecklen Mfg. Individual parts (seen in Figure 4) were electron beam welded together at another local vendor, Sciaky Inc.

The cavity was coarsely tuned by iterations of wire EDM cutting the housing (on both the toroid and dome side) followed by a fit up with indium for a frequency check. The cavity was hand sanded up to 400 grit to remove large surface imperfections.

## INITIAL SURFACE PROCESSING

Following tuning and fabrication, the cavity went through a bulk electropolishing process. This removed the surface layer of the niobium with a target removal of 120 microns. The mirror finish is seen below in Figure 5.

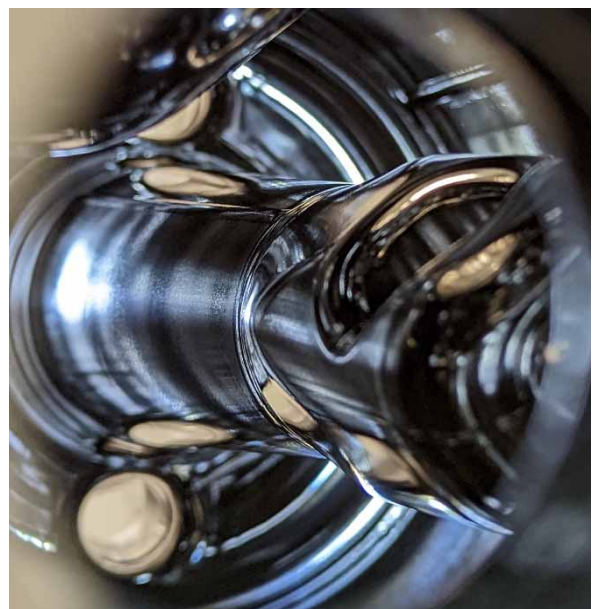

Figure 5. Post-EP cavity surface finish.

## CAVITY COLD TESTING

Before coating, a baseline cold test was required to ensure the bare niobium cavity performed satisfactorily. The baseline requirement was to reach a peak surface magnetic field of 60 mT. Before cold testing, the cavity was ultrasonically cleaned and high pressure rinsed, seen in Figure 6.

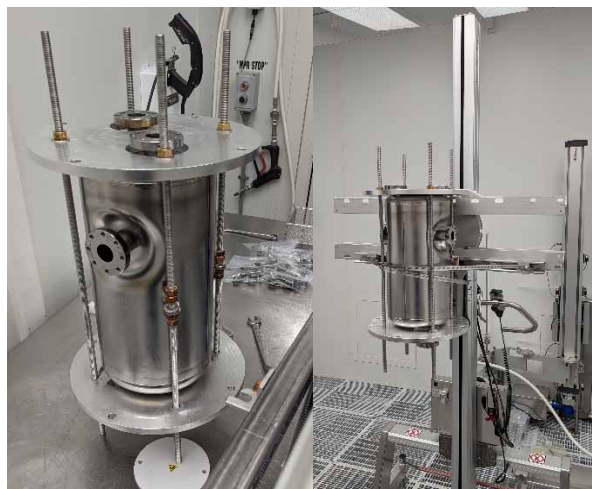

Figure 6. Cavity fixturing used for ultrasonic cleaning and high pressure rinsing.

The cavity was placed in a stainless steel vessel designed for the test cryostat TC3 at Argonne, seen in Figure 7. The vessel fills entirely with liquid helium while allowing RF and vacuum pumping connections to be made. This allowed for active pumping during the testing.

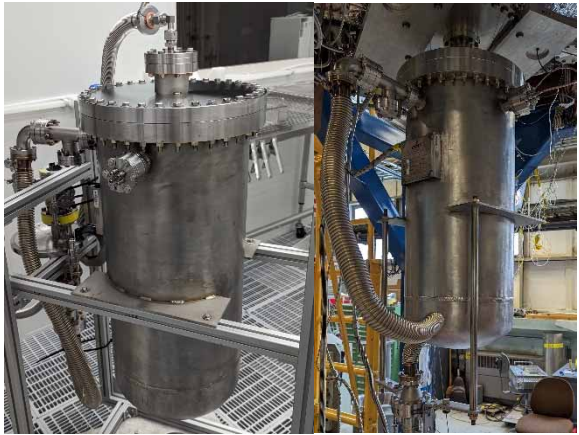

Figure 7. Helium vessel used for cavity cold testing.

Cavity testing saw numerous hurdles with RF connections and liquid helium refrigerator issues, with a final cold test being completed late May.

The initial cold test suffered from RF cabling issues, with coupling becoming exceedingly weak at cryogenic temperatures. This limited the power input and ultimate field level achieved but allowed for a low field Q measurement to be made.

The second cold test was successful but was limited to very low field level by breakdown (Vacc < 10 kV). This was later found to be caused by very low field multipacting, but the field breakdown looked different than multipacting that had been seen before.

This breakdown prompted us to warm and re-clean the cavity and replace a suspicious antenna. The subsequent cooldown suffered identical low field breakdown. This prompted exploration of the next cavity mode, $3\lambda/4$. This HOM of the cavity ran without low field limit and saw conditioning through a multipacting band that looked like multipacting the group had seen before (steady cavity field level for increasing input power).

After running at the $3\lambda/4$ mode, the cavity was once again run at 218 MHz and still saw low field breakdown. However, when driving the cavity with high power without stepping up gradually the low field breakdown was not seen. This is when the breakdown was determined to be very low field multipacting, and the cavity performed as expected at power levels above this multipacting band.

The cavity performed well with a low-field $Q_0$ of $8.2 \times 10^8$, and reached an accelerating gradient of $E_{ACC} = 8.7$ MV/m. The Q curve is given in Figure 8. This corresponded to a $B_{peak} = 59.5$ mT and an $E_{peak} = 49$ MV/m. Some field emission was seen but ultimate field level was limited by RF amplifier power and coupling strength.

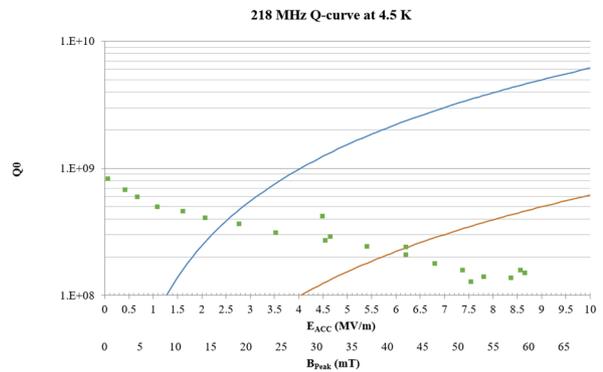

Figure 8. Q curve measurement of 218 MHz cavity.

One of the benefits of the multiple cooldowns performed was to see how cooldown time affected $Q_0$. The initial cooldown (470 minutes) was significantly longer and had a $Q_0$ of $5.7 \times 10^8$. The faster cooldowns were 130 minutes ($Q_0 = 7 \times 10^8$) and 190 minutes ($Q_0 = 8.2 \times 10^8$). We believe this indicates Q disease is present, which is expected with no post-EP bake out. Improvements between the faster cooldowns may also be explained by degaussing performed on the stainless steel vessel, reducing trapped magnetic flux.

Additional cavity performance factors were measured as well, particularly those of microphonics and df/dp (seen in Figure 9). These are informative data points taken with this bare cavity test, but final sensitivities will ultimately depend on final configuration for use, including helium jacketing and attached cryostat features.

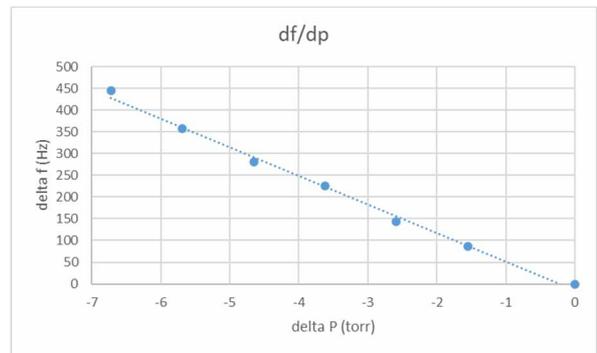

Figure 9. df/dp measurement showing -65.9 Hz/torr sensitivity.

## COATING PROCESS

After successful testing, the cavity was disassembled from the test setup and prepared for coating in the vacuum furnace at Fermilab. This required anodizing the niobium, done with a 30 V DC power supply, which turns the niobium blue (Figure 10). The cavity was ultrasonic cleaned and high pressure rinsed to eliminate contaminants. Additional cavity preparations for coating include fitting up flange covers (to cover NbTi), winding heater coils, and setting up Sn and $SnCl_2$ sources.

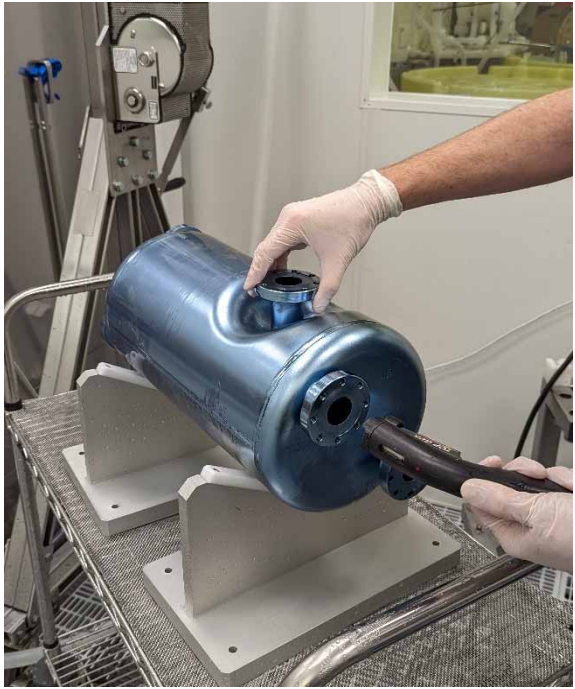

Figure 10. Bare cavity after anodization, a preparation for Nb3Sn coating.

The cavity coating process follows the temperature curve shown in Figure 11. This begins with a nucleation phase, with the temperature held at about 500°C, allowing the $SnCl_2$ to nucleate tin sites on the surface. Coating is then done with the cavity held at 1100°C and the Sn sources at ~1200°C. The tin sources are heated above cavity temperature with molybdenum wire heaters, seen in Figure 12 [3].

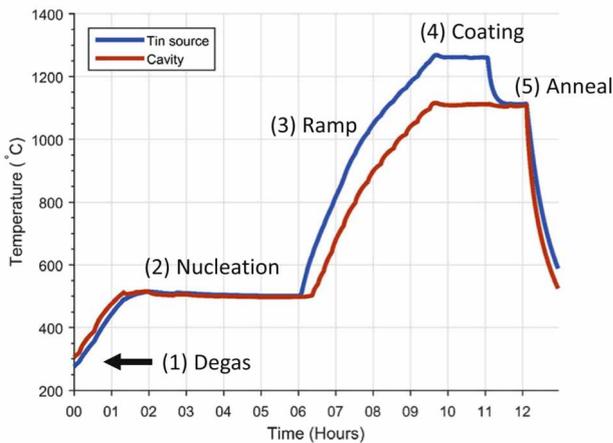

Figure 11. Typical niobium-tin coating process heat cycle.

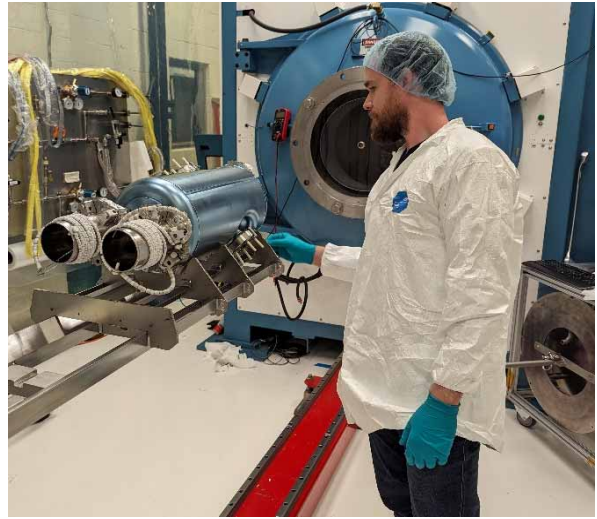

Figure 12. Cavity being prepared for coating in vacuum furnace.

The cavity was successfully coated and awaits a cold test to verify RF performance. The coating process removes the blue anodization and leaves a more matte finish seen in Figure 13.

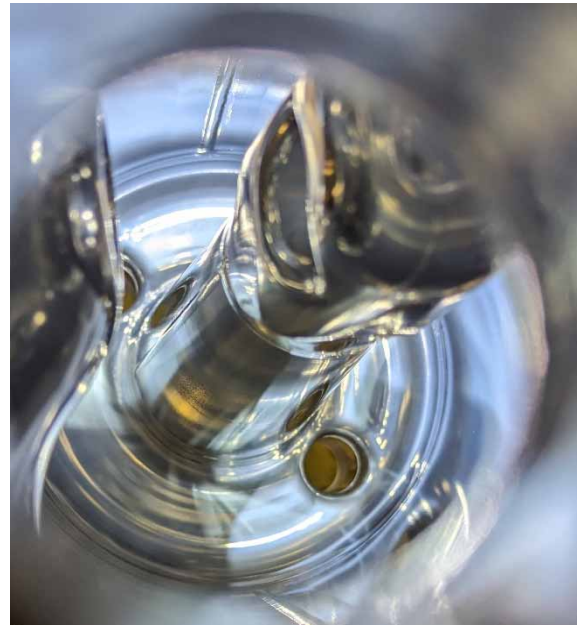

Figure 13. Post-coating cavity surface.

## FUTURE WORK

Planned additional work on this cavity mainly consists of cold testing following surface coating. This includes the normal high pressure rinsing and clean assembly of cavity hardware.

This will be followed by a cold test to verify performance of the cavity post-coating. If it is determined that the Nb3Sn coating was a failure (through visual inspection or from poor performance during testing) additional chemistry will be performed to remove the coated surface before an additional attempt.

Following a demonstration of a high quality factor and surface magnetic field focus will shift to production of a 145 MHz cavity to be used as a re-buncher in ATLAS. This will necessitate building a coarse tuner system, with many approaches being considered. Traditional mechanical tuning may be difficult as there is a material strength difference between Nb and $Nb_3Sn$ that can cause cracking of the SRF surface.

## SUMMARY

A 218 MHz quarter wave cavity has been fabricated and coated with $Nb_3Sn$, the most promising next generation SRF material. Initial bare Nb cavity performance was tested before the coating process, and a planned cold test will verify the coated cavity performance. Successful demonstration of SRF performance will lead to further work fabricating a 145 MHz cavity that will ultimately be installed in ATLAS as a re-buncher.

## ACKNOWLEDGEMENTS

We would like to thank G. Chen for her initial work fabricating this cavity, working with vendors and documenting the process. Additional thank you to our colleagues at Fermilab, particularly Sam Posen and Brad Tennis, who have been critical partners in the project and facilitating coating of this cavity.